\documentclass[conference]{IEEEtran}
\IEEEoverridecommandlockouts

\usepackage{cite}
\usepackage{amsmath,amssymb,amsfonts}
\usepackage{algorithmic}
\usepackage{graphicx}
\usepackage{textcomp}
\usepackage{xcolor}
\usepackage[utf8]{inputenc}
\usepackage{subfigure}
\usepackage{placeins}
\usepackage[ruled,vlined]{algorithm2e}
\def\BibTeX{{\rm B\kern-.05em{\sc i\kern-.025em b}\kern-.08em
    T\kern-.1667em\lower.7ex\hbox{E}\kern-.125emX}}
\begin{document}

\title{The Derivation of The Probability of Error for BPSK, 16-QAM and 64-QAM in Rayleigh Fading Channel: A Unified Approach\\
}

\author{
	\IEEEauthorblockN{1\textsuperscript{st} Amr Abdelbari}
	\IEEEauthorblockA{\textit{Artificial Intelligence Engineering Dept.,} \\
		\textit{AI and Robotics Institute} \\
		\textit{Near East University} \\
		Nicosia, TRNC, Mersin 10, Turkey \\
		amr.abdelbari@neu.edu.tr}
}

\maketitle

\begin{abstract}
Understanding the probability of error is paramount in the design and analysis of digital communication systems, particularly in Rayleigh fading channels where signal impairments are prevalent. This article presents a unified approach for deriving the probability of error formulations applicable to Binary Phase Shift Keying (BPSK), 16-Quadrature Amplitude Modulation (16-QAM), and 64-QAM in Rayleigh fading channels.
This article presents a general approach to derive the probability of error formulations applicable to Binary Phase Shift Keying (BPSK), 16-Quadruple Amplitude Modulation (16-QAM) and 64-QAM in Rayleigh fading channels.
The derivation process encompasses the unique characteristics of each modulation scheme and the statistical properties of Rayleigh fading providing a comprehensive framework to analyze error performance. By establishing a unified formulation, this approach simplifies the analysis and facilitates a deeper understanding of error probability behavior across different modulation schemes.
The derived formulations offer insights into the impact of fading-induced impairments on system performance, allowing for accurate prediction and optimization of communication systems in real-world fading environments. The insights and techniques presented herein serve as a valuable resource for researchers, engineers, and practitioners engaged in the design and optimization of communication systems operating in challenging fading environments.
\end{abstract}


\section{Introduction}\label{Introduction}
In the realm of digital communication systems, the reliable transmission of information is paramount, especially in environments characterized by fading channels. Rayleigh fading, a prevalent phenomenon in wireless communication, introduces random variations in the received signal amplitude and phase due to multipath propagation and environmental factors. Understanding and mitigating the effects of Rayleigh fading are essential for ensuring robust and efficient communication systems \cite{9382012}.

Modulation schemes play a crucial role in shaping the performance of communication systems, offering different trade-offs between spectral efficiency, complexity, and robustness to channel impairments. Among the various modulation techniques, BPSK, 16-QAM, and 64-QAM are widely utilized in practical applications due to their simplicity and efficiency \cite{1306627}.
However, the integration of these modulation schemes into Rayleigh fading channels poses significant challenges, primarily due to the unpredictable nature of fading-induced impairments. Analyzing and quantifying the performance of BPSK, 16-QAM, and 64-QAM in fading channels requires a thorough understanding of the probability of error, which measures the likelihood of incorrect symbol detection at the receiver.
In this paper, we present a unified approach to derive the probability of error formulations for BPSK, 16-QAM, and 64-QAM in Rayleigh fading channels. Our approach encompasses the statistical properties of Rayleigh fading and the modulation characteristics of each scheme, providing a comprehensive framework for analyzing error performance. By establishing a unified formulation, we aim to simplify the analysis process and facilitate a deeper understanding of error behavior across different modulation schemes.

In this paper, we have consistently used specific notation to clarify and standardize the descriptions of statistical distributions and properties. 
i.i.d. stands for independent identically-distributed, indicating that the elements or samples discussed are both independent from each other and share the same probability distribution.
$\sim$ is used to denote statistically distributed as. It precedes the description of the statistical distribution that a variable follows, linking variables to their respective distributions clearly and succinctly. 
$\mathcal{CN}(0, \sigma)$ represents a complex normal vector. This notation specifies that the vector has a zero mean and a variance denoted by $\sigma$. The complex normal distribution, often essential in communications theory, particularly in modeling noise or other random processes, is fundamental for understanding noise characteristics in various signal processing contexts.

\begin{figure}
	\centering
	\includegraphics[width=\linewidth]{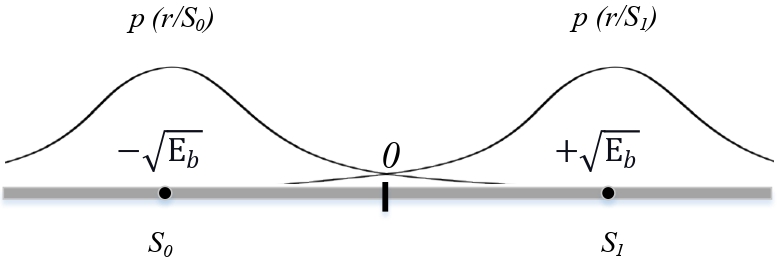}
	\caption{The probability density function PDF for BPSK in AWGN.}
	\label{fig:probabilitydensityfunctionpdfbpskawgn}
\end{figure}

\section{The Calculation of BER for BPSK modulation scheme in AWGN channel}\label{proposed_technique}
The derivation begins by understanding the properties of  BPSK modulation within an Additive White Gaussian Noise (AWGN) channel. BPSK represents binary digits, 1 and 0, through analog levels $+\sqrt{E_b}$ and $-\sqrt{E_b}$ respectively. The key characteristics of AWGN are: additive noise is added to the signal, not multiplied, white noise has a flat spectrum across all frequencies and finally, Gaussian noise values follow a Gaussian distribution, described mathematically as:
\begin{equation}\label{noise_gaussian}
	p(x) = \frac{1}{\sqrt{2\pi \sigma^2}}e^{-\frac{(x - \mu)^2}{2\sigma^2}},
\end{equation}
with $\mu = 0$ and $\sigma^2 = \frac{N_0}{2}$.
The received signal is either $y = s_1 + n$ when bit 1 is transmitted, or $y = s_0 + n$ when bit 0 is transmitted. The conditional probability distribution functions (PDFs) for these scenarios are:
\begin{equation}
	p(y|s_0) = \frac{1}{\sqrt{2\pi \sigma^2}}e^{-\frac{(y +\sqrt{E_b})^2}{2\sigma^2}},
\end{equation}
and
\begin{equation}
	p(y|s_1) = \frac{1}{\sqrt{2\pi \sigma^2}}e^{-\frac{(y - \sqrt{E_b})^2}{2\sigma^2}}.
\end{equation}
Given that both $s_0$ and $s_1$ are equally probable, the optimal decision boundary is at 0. Thus, if $y > 0$, $s_1$ is assumed transmitted; if $y \leq 0$, $s_0$ is assumed transmitted.

\subsubsection{Probability of error given $s_{1}$ was transmitted}
For $s_1$ transmission, the probability of error is determined by the integral of the tail of the Gaussian distribution extending from $-\infty$ to 0:
\begin{equation}\label{BPSK_BER_derivation}
	p(e|s_1) = \frac{1}{\sqrt{2\pi \sigma^2}} \int_{-\infty}^0 e^{-\frac{(y - \sqrt{E_b})^2}{2\sigma^2}} dy.
\end{equation}
Using a substitution where $z^{2} = \dfrac{(y - \sqrt{E_{b}})^{2}}{N_{0}}$ and $N_{0} = 2\sigma^{2}$. Thus, $z = \dfrac{y - \sqrt{E_{b}}}{\sqrt{N_{0}}}$, $dz = \dfrac{1}{\sqrt{N_{0}}}dy$, and $dy = \sqrt{N_{0}}dz$. The integration limits will be as follows: $y = 0 \Rightarrow z = - \sqrt{\dfrac{E_{b}}{N_{0}}}$ and $y = \infty \Rightarrow z = \infty$.
Equation (\ref{BPSK_BER_derivation}) simplifies to:
\begin{equation}
	\begin{split}
		p(e|s_{1}) = & \dfrac{\sqrt{N_{0}}}{\sqrt{\pi N_{0}}} \int_{\sqrt{\frac{E_{b}}{N_{0}}}}^{\infty} e^{-z^{2}} dz  \\
		= & \dfrac{2}{2\sqrt{\pi}} \int_{\sqrt{\frac{E_{b}}{N_{0}}}}^{\infty} e^{-z^{2}} dz  \\
		= & \dfrac{1}{2} erfc(\sqrt{\frac{E_{b}}{N_{0}}}),
	\end{split}
\end{equation}
where 
\begin{equation}
		erfc(x) =  \dfrac{2}{\sqrt{\pi}} \int_{x}^{\infty} e^{-x^{2}} dx.
\end{equation}

\subsubsection{Probability of error given $s_{0}$ was transmitted}
Similarly, the error probability given $s_0$ was transmitted is also:
\begin{equation}
	\begin{split}
		p(e|s_{0}) = & \dfrac{1}{\sqrt{2 \pi \sigma^{2}}} \int_{0}^{\infty} e^{\dfrac{-(y + \sqrt{E_{b}})^{2}}{2\sigma^{2}}} dy  \\
		= & \dfrac{1}{\sqrt{\pi}} \int_{\sqrt{\frac{E_{b}}{N_{0}}}}^{\infty} e^{-z^{2}} dy  \\
		= & \dfrac{1}{2} erfc(\sqrt{\frac{E_{b}}{N_{0}}}).
	\end{split}
\end{equation}

\subsubsection{Total probability of bit error}
Generally, the total probability of bit error can be calculated as follows:
\begin{equation}
	P_{b} = p(s_{1})p(e|s_{1}) + p(s_{0})p(e|s_{0}).
\end{equation}
Considering equal probabilities for $s_0$ and $s_1$, i.e.  $p(s_{0}) = p(s_{1}) = \frac{1}{2}$, the total probability of bit error is: 
\begin{equation}\label{BER_BPSK}
	P_{b} = \dfrac{1}{2} erfc\big(\sqrt{\frac{E_{b}}{N_{0}}}\big).
\end{equation}

\section{The Calculation of BER for BPSK modulation scheme in Rayleigh Channel}
When extending the analysis to a Rayleigh fading channel, the channel response $h = h_{re} + jh_{im},$ is modeled as a circularly symmetric complex Gaussian random variable, where the magnitude $|h|$ follows a Rayleigh distribution. This channel model reflects an environment with numerous reflectors.
The real $h_{re}$ and imaginary $h_{im}$ parts are i.i.d. with mean 0 and variance $\sigma_{2}$.
The probability density of the magnitude $|h|$ is,
\begin{equation}
	p(h) = \dfrac{h}{\sqrt{\sigma^{2}}}e^{\dfrac{-h^{2}}{2\sigma^{2}}}.
\end{equation}

\subsection{System model}
The received signal in a Rayleigh fading channel is:
\begin{equation}
	y = hx + n,
\end{equation}
where $x$ represents the transmitted symbol ($\pm 1$) and $n$ is the AWGN. The channel is assumed to be flat fading and randomly varying in time, with the channel $h$ known at the receiver, allowing for straightforward equalization.
The equalization technique employed in this scenario involves a straightforward division of the received signal $y$ by the apriori known channel coefficient $h$.the estimated transmitted symbol after equalization is
\begin{equation}
	\hat{y} = \frac{y}{h} = \frac{hx + n}{h} = x + \frac{n}{h} = x + \tilde{n},
\end{equation}
where $\tilde{n} = \frac{n}{h}$ denotes the scaled noise, resulting from the division of the original noise by the channel coefficient.

\subsection{Bit Error Rate}
In a Rayleigh fading scenario, the effective bit energy to noise ratio becomes $\frac{|h|^2 E_b}{N_0}$. The conditional BER given the channel $h$ is:
\begin{equation}\label{}
	P_{b|h} = \dfrac{1}{2} erfc\big(\sqrt{\frac{|h|^{2}E_{b}}{N_{0}}}\big) = \dfrac{1}{2} erfc\big(\sqrt{\gamma}\big),
\end{equation}
where $\gamma = \frac{|h|^{2}E_{b}}{N_{0}}$.
The unconditional BER incorporates the distribution of $|h|^2$, which affects the overall performance due to the fading nature of the channel \cite{Proakis2007}.

It is known that if $|h|$ is a Rayleigh distributed random variable, then $|h|^{2}$ is chi-square distributed with two degrees of freedom. since $|h|^{2}$ is chi square distributed, $\gamma$ is also chi square distributed. The probability density function of $\gamma$ is,
\begin{equation}
	\begin{split}
		p(\gamma) = & \frac{1}{E_{b}/N_{0}}e^{\frac{-\gamma}{E_{b}/N_{0}}}, ~ \gamma \ge 0 \\
		= & \frac{1}{\bar{\gamma}}e^{\frac{-\gamma}{\bar{\gamma}}} \\
	\end{split}
\end{equation}
where $\bar{\gamma} = E_{b}/N_{0}$.

To determine the overall error probability in a Rayleigh fading channel, we integrate the conditional probability of error over all possible values of the channel power $\gamma$. The probability of bit error $P_b$ in a Rayleigh fading channel is computed by 
\begin{equation}
	P_{b} =  \int_{0}^{\infty}  \dfrac{1}{2} erfc\big(\sqrt{\gamma}\big)p(\gamma) d\gamma.
\end{equation}
To evaluate the integral, we leverage several mathematical properties and functions detailed in the Appendix \ref{Common_used_Probabilities}, and other common mathematical resources. Specifically, the derivatives of the error function (erf) and its complement (erfc) are used to facilitate the integration by parts, a technique employed to simplify the evaluation of integrals involving product terms.
The derivative of the complementary error function with respect to a variable $x$ is given by:
\begin{equation}
	\dfrac{d}{dx} erfc\big(\sqrt{x}\big) = -\dfrac{1}{\sqrt{\pi }} e^{-x^{2}}x^{-\frac{1}{2}}
\end{equation}
\begin{equation}
	\dfrac{d}{dx} erf\big(\sqrt{x}\big) = \dfrac{1}{\sqrt{\pi }} e^{-x^{2}}x^{-\frac{1}{2}}
\end{equation}
These derivatives are crucial for handling the exponential terms that arise during the integration by parts process. The method of integration by parts is applied to solve the integral involving $erfc\big(\sqrt{\gamma}\big)$ and the exponential decay of $\gamma$ as described by its probability distribution function (PDF). The integral setup typically follows the formula ( $\int u dv = uv - \int v du$ ),
where choosing appropriate functions for $u$ and $dv$ is essential to simplify the integration process. For our case, $u$ might be chosen as $erfc(\sqrt{x})$, $du = -\dfrac{1}{\sqrt{x\pi}} e^{-x} dx$ and $dv$ as $e^{-\frac{x}{\alpha}}dx$, $v = -\alpha e^{-\frac{x}{\alpha}}$ allowing us to express the integral in terms that can be more straightforwardly evaluated or approximated.
\begin{equation}\label{helping_erfc_intergral}
	\begin{split}
	\int erfc\big(\sqrt{x}\big)e^{-\frac{x}{\alpha}}dx = & -\alpha erfc(\sqrt{x})e^{-\frac{x}{\alpha}} - \\
	& \int (-\alpha)e^{-\frac{x}{\alpha}} (-\frac{1}{\sqrt{x\pi}})e^{-x} dx \\
		\end{split}
\end{equation}
The second part can be solved as follows:
\begin{equation}
	\begin{split}
		\frac{\alpha}{\sqrt{\pi}}\int e^{-\frac{x}{\alpha}}e^{-x} x^{-\frac{1}{2}} dx  = & \frac{\alpha}{\sqrt{\pi}}\int e^{-\frac{x(\alpha + 1)}{\alpha}} x^{-\frac{1}{2}} dx  \\
	\end{split}
\end{equation}
By taking $u = \frac{x(\alpha + 1)}{\alpha}$, $x = \frac{(\alpha)}{\alpha+1}u$, and $dx = \frac{(\alpha)}{\alpha+1}du$, the previous integral reduces to:
\begin{equation}
	\begin{split}
		\frac{\alpha}{\sqrt{\pi}}\int e^{-\frac{x}{\alpha}}e^{-x} x^{-\frac{1}{2}} dx  = & \frac{\alpha}{\sqrt{\pi}}\int e^{-u} (\frac{\alpha}{\alpha+1})^{-\frac{1}{2}} u^{-\frac{1}{2}}  (\frac{\alpha}{\alpha+1})du  \\
		= & \frac{\alpha}{\sqrt{\pi}}(\frac{\alpha}{\alpha+1})^{\frac{1}{2}}\int e^{-u}  u^{-\frac{1}{2}}  du  \\
	\end{split}
\end{equation}
By taking $t = \sqrt{u}$, $t^{2} = u$, and $du =2tdt$, the previous integral reduces to:
\begin{equation}
	\begin{split}
		\frac{\alpha}{\sqrt{\pi}}(\frac{\alpha}{\alpha+1})^{\frac{1}{2}}\int e^{-u}  u^{-\frac{1}{2}} du  = & \frac{\alpha}{\sqrt{\pi}}(\frac{\alpha}{\alpha+1})^{\frac{1}{2}}\int e^{-t^{2}}  t^{-\frac{2}{2}} 2 t dt   \\
		= & \frac{2\alpha}{\sqrt{\pi}}(\frac{\alpha}{\alpha+1})^{\frac{1}{2}}\int e^{-t^{2}} dt   \\
		= & \alpha(\frac{\alpha}{\alpha+1})^{\frac{1}{2}} erf(\sqrt{u}) \\
		= & \alpha(\frac{\alpha}{\alpha+1})^{\frac{1}{2}} erf(\sqrt{\frac{(\alpha + 1)}{\alpha}x}) \\
	\end{split}
\end{equation}
By replacing the last part with the second part in (\ref{helping_erfc_intergral}):
\begin{equation}\label{helping_erfc_intergral_2}
	\begin{split}
		\int erfc\big(\sqrt{x}\big)e^{-\frac{x}{\alpha}}dx = & -\alpha erfc(\sqrt{x})e^{-\frac{x}{\alpha}} - \\
		& \alpha(\frac{\alpha}{\alpha+1})^{\frac{1}{2}} erf(\sqrt{\frac{(\alpha + 1)}{\alpha}x})\\
	\end{split}
\end{equation}
With the preceding analysis, deriving the BER for BPSK modulation in a Rayleigh fading channel becomes straightforward.  In (\ref{helping_erfc_intergral_2}), let $\alpha = \bar{\gamma}$ and $x = \gamma$:
\begin{equation}
	\begin{split}
	P_{b} =  &\int_{0}^{\infty}  \dfrac{1}{2\bar{\gamma}} erfc\big(\sqrt{\gamma}\big)e^{\frac{-\gamma}{\bar{\gamma}}} d\gamma \\
	= & \dfrac{1}{2\bar{\gamma}} \big[ -\bar{\gamma} erfc(\sqrt{\gamma})e^{-\frac{\gamma}{\bar{\gamma}}} - \\
		& \bar{\gamma}\sqrt{\frac{\bar{\gamma}}{\bar{\gamma}+1}} erf(\sqrt{\frac{\bar{\gamma} + 1}{\bar{\gamma}}\gamma})   \big]_{0}^{\infty} \\
	= & \dfrac{1}{2\bar{\gamma}} \big[ \bar{\gamma} erfc(\sqrt{\gamma})e^{-\frac{\gamma}{\bar{\gamma}}} + \\
		& \bar{\gamma}\sqrt{\frac{\bar{\gamma}}{\bar{\gamma}+1}} erf(\sqrt{\frac{\bar{\gamma} + 1}{\bar{\gamma}}\gamma})   \big]_{\infty}^{0} \\
	\end{split}
\end{equation}
Taking $erfc(0) = 1$, $erfc(\infty) = 0$, $erf(0) = 0$ and $erf(\infty) = 1$:

\begin{equation}
	\begin{split}
		P_{b}
		= & \dfrac{1}{2\bar{\gamma}} \big[ (\bar{\gamma} + 0) -  (0 + \bar{\gamma}\sqrt{\frac{\bar{\gamma}}{\bar{\gamma}+1}} )\big] \\
		= & \dfrac{\bar{\gamma}}{2\bar{\gamma}} \big[ (1 - \sqrt{\frac{\bar{\gamma}}{\bar{\gamma}+1}} )\big] \\
		= & \dfrac{1}{2} \big[ (1 - \sqrt{\frac{E_{b}/N_{0}}{E_{b}/N_{0}+1}} )\big] \\
	\end{split}
\end{equation}

\begin{figure}
	\centering
	\includegraphics[width=\linewidth]{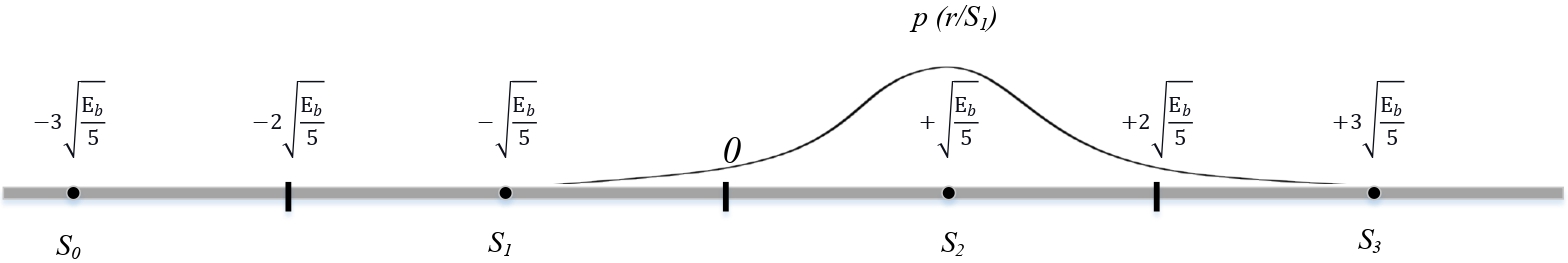}
	\caption{Figure: Constellation plot for 4 PAM modulation.}
	\label{fig:4pampdfalphabet1}
\end{figure}

\section{The Calculation of BER for 4-PAM Modulation Scheme in AWGN Channel}

Consider the symbols used in a 4-PAM system, which are defined as $\alpha_{4-QAM} = {\pm 1, \pm 3}$.
The average energy of the PAM constellation, assuming each symbol is equally likely to occur, can be calculated as follows:
\begin{equation}
	E_{4PAM} = E\big[ |\alpha_{4PAM}|^{2} \big] = \frac{1}{4}\sum[1^2 + 3^2] = 5
\end{equation}
This expression considers the squared amplitude of each symbol, reflecting the energy per symbol in the constellation \cite{BarryLeeMesserschmitt04_DigitalCommunication}.
A normalized plot of the 4-PAM signal's constellation can be visualized as depicted in Fig. \ref{fig:4pampdfalphabet1}. It is assumed that the additive noise, $n$, adhering to the Gaussian probability distribution function, impacts the received signals, as described in equation (\ref{noise_gaussian}).
The received signal for a transmitted symbol $s_i$ can be modeled by:
\begin{equation}
	y = s_{i} + n, \quad i = 1, 2, 3, 4
\end{equation}
Focusing on the scenario where $s_{3}$ is transmitted, the conditional PDF for the received signal $y$ is:
\begin{equation}
	p(y | s_{3}) = \frac{1}{\sqrt{\pi N_{0}}} e^{- \frac{(y - 3\sqrt{\frac{E_{s}}{5}})^{2}}{N_{0}}}
\end{equation}
This scenario assumes the detection threshold between the signals for +1 and +3 is set at $+2\sqrt{\frac{E_{s}}{5}}$. Consequently, if the received signal $y$ exceeds this threshold, the receiver interprets that $s_{3}$ was transmitted.

Using this threshold, the probability of error when $s_{3}$ is transmitted can be calculated to ensure accurate detection and error assessment in noisy environments by
\begin{equation}\label{4PAM_BER_derivation}
	p(e|s_{3}) = \dfrac{1}{\sqrt{\pi N_{0}}} \int_{-\infty}^{+2\sqrt{\frac{E_{s}}{5}}} e^{- \dfrac{(y - 3\sqrt{\frac{E_{s}}{5}})^{2}}{N_{0}}} dy 
\end{equation}
By taking $z^{2} = \dfrac{(y - 3\sqrt{\frac{E_{s}}{5}})^{2}}{N_{0}}$. Thus, $z = \dfrac{y - 3\sqrt{\frac{E_{s}}{5}}}{\sqrt{N_{0}}}$, $dz = \dfrac{1}{\sqrt{N_{0}}}dy$, and $dy = \sqrt{N_{0}}dz$. The integration limits will be as follows: $y = +2\sqrt{\frac{E_{s}}{5}} \Rightarrow z = - \sqrt{\dfrac{E_{s}}{5N_{0}}}$ and $y = -\infty \Rightarrow z = -\infty$.
Therefore, (\ref{4PAM_BER_derivation})will be:
\begin{equation}
	\begin{split}
		p(e|s_{3}) = & \dfrac{\sqrt{N_{0}}}{\sqrt{\pi N_{0}}} \int_{\sqrt{\frac{E_{s}}{5N_{0}}}}^{\infty} e^{-z^{2}} dz  \\
		= & \dfrac{2}{2\sqrt{\pi}} \int_{\sqrt{\frac{E_{s}}{5N_{0}}}}^{\infty} e^{-z^{2}} dz  \\
		= & \dfrac{1}{2} erfc(\sqrt{\frac{E_{s}}{5N_{0}}}).
	\end{split}
\end{equation}
 \cite{Proakis2007}
Given the symmetrical nature of the 4-PAM constellation, particularly regarding the symbols +3 and -3, it is intuitively reasonable to conclude that the probability of error when $s_{3}$ (representing +3) is transmitted mirrors that when -3 is transmitted. This symmetry in the constellation layout implies that the error characteristics for these signals are identical due to their equidistant spacing from their respective decision boundaries in the presence of Gaussian noise.

Focusing now on the symbol $s_{2}$, which typically represents one of the intermediate values in a PAM constellation, the probability of error when $s_{2}$ is transmitted can be calculated by
\begin{equation}\label{4PAM_BER_derivation_s_2}
	\begin{split}
			p(e|s_{2}) = & \dfrac{1}{\sqrt{\pi N_{0}}} \int_{-\infty}^{0} e^{- \dfrac{(y - \sqrt{\frac{E_{s}}{5}})^{2}}{N_{0}}} dy  + \\
				& \dfrac{1}{\sqrt{\pi N_{0}}} \int_{+2\sqrt{\frac{E_{s}}{5}}}^{\infty} e^{- \dfrac{(y - \sqrt{\frac{E_{s}}{5}})^{2}}{N_{0}}} dy 
	\end{split}
\end{equation}
For the first two part of (\ref{4PAM_BER_derivation_s_2}), take $z^{2} = \dfrac{(y - \sqrt{\frac{E_{s}}{5}})^{2}}{N_{0}}$. Thus, $z = \dfrac{y - \sqrt{\frac{E_{s}}{5}}}{\sqrt{N_{0}}}$, $dz = \dfrac{1}{\sqrt{N_{0}}}dy$, and $dy = \sqrt{N_{0}}dz$. The integration limits of the first integral will be $y = 0 \Rightarrow z = - \sqrt{\dfrac{E_{s}}{5N_{0}}}$ and $y = -\infty \Rightarrow z = -\infty$. The integration limits of the second integral will be $y = 2\sqrt{\dfrac{E_{s}}{5N_{0}}} \Rightarrow z = \sqrt{\dfrac{E_{s}}{5N_{0}}}$ and $y = \infty \Rightarrow z = \infty$.
Therefore, (\ref{4PAM_BER_derivation_s_2}) will be:
\begin{equation}
	\begin{split}
		p(e|s_{2}) = & \dfrac{1}{\sqrt{\pi}} \int_{- \infty}^{- \sqrt{\frac{E_{s}}{5N_{0}}}} e^{-z^{2}} dz +  \\ 
		 & \dfrac{1}{\sqrt{\pi}} \int_{\sqrt{\frac{E_{s}}{5N_{0}}}}^{\infty} e^{-z^{2}} dz  \\
		= & \dfrac{2}{\sqrt{\pi}} \int_{\sqrt{\frac{E_{s}}{5N_{0}}}}^{\infty} e^{-z^{2}} dz  \\
		= & erfc(\sqrt{\frac{E_{s}}{5N_{0}}}).
	\end{split}
\end{equation}
Considering the symmetrical arrangement of the +1 and -1 symbols in the 4-PAM constellation, it logically follows that the probability of error when transmitting $s_{1}$ (representing +1) is identical to that when transmitting $s_{2}$ ( representing -1).
Since each symbol in the constellation is assumed to be equally likely, the total probability of symbol error across the constellation can be determined by 
\begin{equation}
	P_{s} = p(s_{0})p(e|s_{0}) + p(s_{1})p(e|s_{1}) + p(s_{2})p(e|s_{2}) + p(s_{3})p(e|s_{3})
\end{equation}
Given that $s_{0}$, $s_{1}$, $s_{2}$  and $s_{3}$ are equally probable  i.e.  $p(s_{0}) = p(s_{1}) = p(s_{2}) = p(s_{3}) = \frac{1}{4}$, the symbol error probability is,
\begin{equation}\label{SER_4PAM}
	P_{s} = \dfrac{1}{4} \big[ erfc\big(\sqrt{\frac{E_{s}}{5N_{0}}}\big) +2erfc\big(\sqrt{\frac{E_{s}}{5N_{0}}}\big) \big]  = \dfrac{3}{4}erfc\big(\sqrt{\frac{E_{s}}{5N_{0}}}\big)
\end{equation}
Since $E_{s}/N_{0} = 2E_{b}/N_{0}$ and BER = $\frac{1}{log_2(M)}$ SER, the bit error probability is \cite{10.1002/9781119264422.app1}:
\begin{equation}\label{BER_4PAM}
	P_{b} = \dfrac{3}{8}erfc\big(\sqrt{\frac{2E_{b}}{5N_{0}}}\big) =  \dfrac{3}{4}Q\big(\sqrt{\frac{4E_{b}}{5N_{0}}}\big) 
\end{equation}
To be expressed in terms of distance between each successive symbols $d = 2\sqrt{\frac{E_{s}}{N_{0}}}$.
\begin{equation}\label{}
	P_{b} = \dfrac{3}{8}erfc\big(\frac{d}{2\sqrt{2}\sigma}\big)
\end{equation}
or generally for any M-PAM \cite{1ab41e9ebd0643a086a0c08b8481b174}:
\begin{equation}\label{BER_MPAM}
	P_{b} = \dfrac{M-1}{Mlog_{2}(M)}erfc\big(\frac{d}{2\sqrt{2}\sigma}\big) =  \dfrac{2(M-1)}{Mlog_{2}(M)}Q\big(\frac{d}{2}\sqrt{\frac{2}{N_{0}}}\big)
\end{equation}

\begin{figure}
	\centering
	\includegraphics[width=0.7\linewidth]{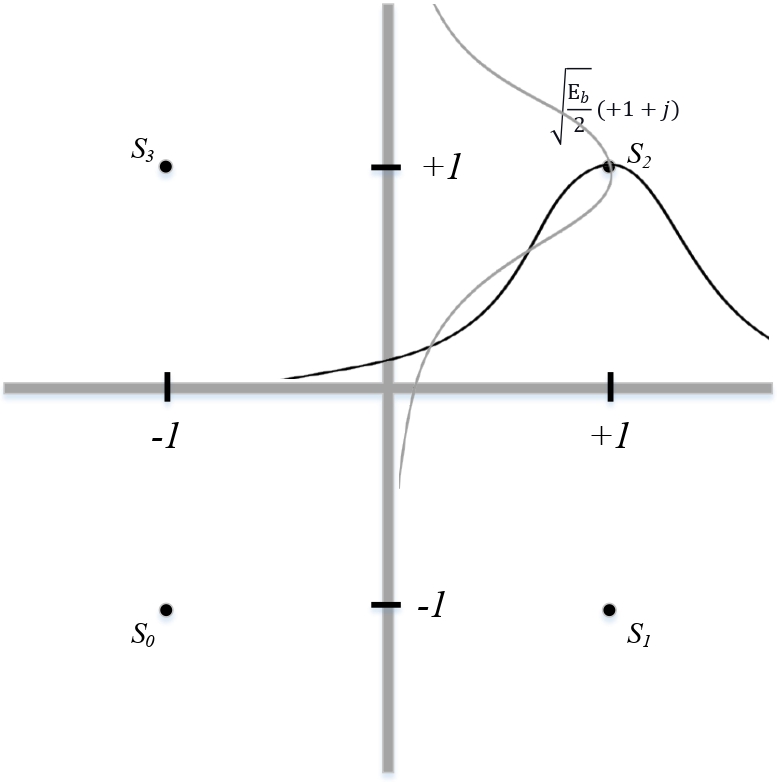}
	\caption{Constellation plot for QPSK (4-QAM) constellation.}
	\label{fig:4qampdf}
\end{figure}

\section{The Calculation of BER for 4-QAM Modulation Scheme in AWGN Channel}

Consider that the alphabets used for a QPSK (4-QAM) is $\alpha_{4QAM} = { \pm 1 \pm1j}$. The scaling factor of $\sqrt{\frac{E_s}{2}}$ is utilized to normalize the average energy of the transmitted symbols to 1. This normalization is based on the assumption that all points in the constellation are equally probable. Such scaling ensures that the total power of the constellation adheres to the system's energy constraints, thereby optimizing the transmission power across different communication scenarios.
It is assumed that the additive noise, denoted by $n$, adheres to the Gaussian probability distribution function, as specified in (\ref{noise_gaussian}).

For the symbol $s_2$, the conditional PDF when $s_2$ is transmitted is given by:
\begin{equation}
	p(y | s_{2}) = \frac{1}{\sqrt{N_{0}\pi}} e^{- \frac{(y - \sqrt{\frac{E_s}{2}})^{2}}{N_{0}}}.
\end{equation}
This equation indicates how the received symbol $y$ is distributed around the transmitted symbol $s_2$ under Gaussian noise.
As illustrated in Fig. \ref{fig:4qampdf}, correct decoding of symbol $s_2$ is contingent upon $y$ landing within a specific hashed region. The probability that the symbol $s_2$ is correctly decoded, considering both the real and imaginary components of $y$, is:
\begin{equation}
	p(c|s_{2}) = p(\Re_{y} > 0 | s_{2}) \cdot p(\Im_{y} > 0 | s_{2})
\end{equation}
This probability effectively represents the likelihood that the real component of $y$ is greater than 0 given that $s_{2}$ was transmitted.
The probability that the real component of the received signal $y$ is greater than 0, given $s_{2}$was transmitted, is critical for successful demodulation and is given by:
\begin{equation}
	\begin{split}
		p(\Re_{y}> 0  | s_{2}) = & 1 - \dfrac{1}{\sqrt{N_{0}\pi}} \int_{-\infty}^{0}e^{- \dfrac{(y - \sqrt{\frac{E_{s}}{2}})^{2}}{N_{0}}}dy \\
		= & 1 - erfc\big(\sqrt{\frac{E_{s}}{2N_{0}}}\big)
	\end{split}	
\end{equation}
This result is derived by calculating the area under the Gaussian curve from negative infinity to zero, subtracted from 1, which indicates the cumulative probability from zero to positive infinity.
Similarly, the probability that the imaginary component of  $y$ is greater than 0, given $s_{2}$ was transmitted is calculated as:
\begin{equation}
	\begin{split}
		p(\Im_{y}> 0  | s_{2}) = & 1 - \dfrac{1}{\sqrt{N_{0}\pi}} \int_{-\infty}^{0}e^{- \dfrac{(y - \sqrt{\frac{E_{s}}{2}})^{2}}{N_{0}}}dy \\
		= & 1 - erfc\big(\sqrt{\frac{E_{s}}{2N_{0}}}\big).
	\end{split}	
\end{equation}
The overall probability that $s_{2}$ is correctly decoded, which considers both the real and imaginary components, is then given by:
\begin{equation}
	\begin{split}
		p(c|s_{2})	= & \big[1 - erfc\big(\sqrt{\frac{E_{s}}{2N_{0}}}\big) \big]^{2} \\
		= & 1 - \dfrac{2}{2}erfc\big(\sqrt{\frac{E_{s}}{2N_{0}}}\big) + \dfrac{1}{4}erfc\big(\sqrt{\frac{E_{s}}{2N_{0}}}\big)^{2} \\
		= & 1 - erfc\big(\sqrt{\frac{E_{s}}{2N_{0}}}\big) + \dfrac{1}{4}erfc\big(\sqrt{\frac{E_{s}}{2N_{0}}}\big)^{2} \\
	\end{split}	
\end{equation}
The probability that the symbol $s_{2}$ is in error, i.e., at least one component is incorrectly decoded, is thus:
\begin{equation}
	\begin{split}
		P_{QPSK} = &1- p(c|s_{2})	\\
		= & 1- \big[ 1 - erfc\big(\sqrt{\frac{E_{s}}{2N_{0}}}\big) + \dfrac{1}{4}erfc\big(\sqrt{\frac{E_{s}}{2N_{0}}}\big)^{2} \big] \\
		= & erfc\big(\sqrt{\frac{E_{s}}{2N_{0}}}\big) - \dfrac{1}{4}erfc\big(\sqrt{\frac{E_{s}}{2N_{0}}}\big)^{2}  \\
	\end{split}	
\end{equation}
For high values of the signal-to-noise ratio \(\frac{E_{s}}{2N_{0}}\), where the system approaches an ideal state with minimal noise, the second term becomes negligible, allowing the error probability to be approximated simply as:
\begin{equation}
	P_{QPSK} \approx erfc\big(\sqrt{\frac{E_{s}}{2N_{0}}}\big) = 2Q\big(\sqrt{\frac{E_{s}}{N_{0}}}\big)
\end{equation}
This approximation highlights that as the SNR increases, the system's performance significantly improves, reducing the error rate and enhancing reliable communication \cite{BarryLeeMesserschmitt04_DigitalCommunication}.

\begin{figure}
	\centering
	\includegraphics[width=0.7\linewidth]{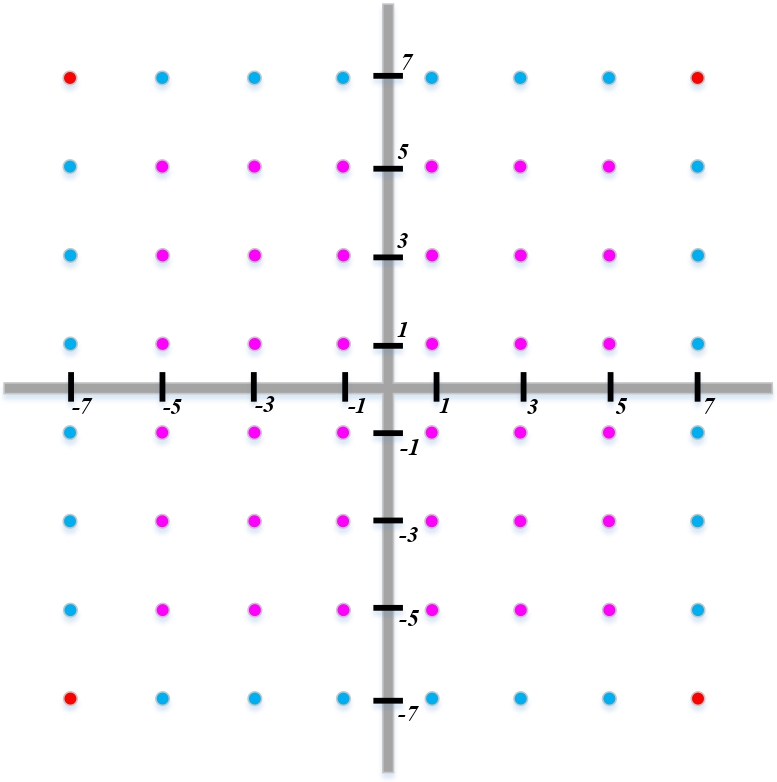}
	\caption{Constellation plot for 64-QAM modulation (without the scaling factor of $\frac{1}{\sqrt{42}}$).}
	\label{fig:64qamconstellationplot}
\end{figure}

\section{The Calculation of BER for M-QAM Modulation Scheme in AWGN Channel}

The M-QAM constellation is defined based on $M = 2^b$, where $b$ represents the number of bits per symbol. For practical and analytical convenience, $b$ is often chosen to be even due to the following advantages. First, half of the bits are mapped onto the real axis and the other half onto the imaginary axis. This division results in two independent $b/2$-level PAM signals, simplifying the design of the mapping scheme.
Second, decoding processes can be independently applied to the real and imaginary components of the signal, which streamlines the receiver architecture.
It is important to note, however, that this square constellation layout is not always the most efficient in terms of signal-to-noise ratio performance.

For an even $b$ within an M-QAM system, the constellation points are typically given by:
\begin{equation}
	\alpha_{MQAM} = { \pm (2m-1) \pm (2m-1)j}, ~ m \in {1, 2, \dots, \frac{\sqrt{M}}{2}}.
\end{equation}
For instance, a 64-QAM constellation, represented by ($M = 64$) constellation, $m \in {1, 2, 3, 4}$.
\begin{gather}\label{alpha_64QAM}
	\resizebox{0.45\textwidth}{!}{$
		\begin{split}
			\alpha_{64QAM} = \begin{bmatrix}
				(\pm 7 \pm 7j) &  (\pm 7 \pm 5j) &  (\pm 7 \pm 3j) &  (\pm 7 \pm 1j)\\
				(\pm 5 \pm 7j)&  (\pm 5 \pm 5j) &   (\pm 5 \pm 3j) &  (\pm 5 \pm 1j) \\
				(\pm 3 \pm 7j) &  (\pm 3 \pm 5j)&  (\pm 3 \pm 3j) &  (\pm 3 \pm 1j) \\
				(\pm 1 \pm 7j)&  (\pm 1 \pm 5j) &  (\pm 1 \pm 3j) &  (\pm 1 \pm 1j) \\
			\end{bmatrix}.
		\end{split}
		$}
\end{gather}

To calculate the average energy across the constellation, calculate the cumulative energy of individual symbols.
\begin{equation}
	E_{4\alpha} = \sum_{m = 1}^{\frac{\sqrt{M}}{2}}\big[ (2m-1) + j(2m-1) \big]^{2}  = \frac{\sqrt{M}}{2}(M-1).
\end{equation}
Each symbol appears $2\sqrt{M}$ times within the constellation. The average energy is then:
\begin{equation}
	E_{MQAM} =  \frac{2\sqrt{M}}{M}E_{\alpha} = \frac{2\sqrt{M}}{M}\frac{\sqrt{M}}{3} (M - 1) = \frac{2}{3}(M - 1)
\end{equation}
For example, for 64-QAM, the average energy is
\begin{equation}
	E_{64-QAM} = \frac{2}{3}(64 - 1) = 42,
\end{equation}
and for 16-QAM,
\begin{equation}
	E_{16-QAM} = \frac{2}{3}(16 - 1) = 10.
\end{equation}
For specific constellations like 64-QAM and 16-QAM, the respective average energies $\frac{1}{\sqrt{10}}$, $\frac{1}{\sqrt{42}}$  can be calculated and used to normalize the transmission power \cite{BarryLeeMesserschmitt04_DigitalCommunication}.

To analyze the symbol error rate, especially within the context of a 64-QAM system and then extended to general M-QAM formats. Fig. \ref{fig:64qamconstellationplot} illustrates the differentiation of constellation points. The corner points (in red) always four points $N_{corner} = 4$.
The inside points (in magenta) computed as $N_{corner} = (\sqrt{M}-2)^2$.
Last points are edge points which are neither corners nor center (in blue). The number of these points is calculated as $N_{side} = 4(\sqrt{M}-2) $. For 64-QAM it is 24 points.

Assuming the received symbol $y$ is affected by $n$ the AWGN in (\ref{noise_gaussian}), the received value is:
\begin{equation}
	y = k\sqrt{E_{s}} + n.
\end{equation}
Here, $k = \sqrt{\dfrac{1}{\frac{2}{3}(M - 1)}}$ acts as the normalizing factor based on the constellation's average energy, and $E_s$ is the symbol energy.

\subsubsection{For constellation points located inside the M-QAM grid}, particularly for those not on the edges or corners, the conditional PDF for the received signal $y$ , given that the symbol ${+k\sqrt{E_{s}}, +k\sqrt{E_{s}}}$ is transmitted, is modeled as:
\begin{equation}
	p(y | inside) = \dfrac{1}{\sqrt{N_{0}\pi}} e^{- \dfrac{(y - k\sqrt{E_{s}})^{2}}{N_{0}}}
\end{equation}
Correct decoding of the symbol ($I=+1,Q=+1$)is contingent upon the real $y ~(\Re_{y})$and imaginary $y ~(\Im_{y})$ components of the received signal $y$ falling within specific bounds:
\begin{equation}
	\resizebox{0.45\textwidth}{!}{$
		p(c| inside) =  p(\Re_{y}> 0, \Re_{y} \le  2k\sqrt{E_{s}} | +1) 
		p(\Im_{y} > 0, \Im_{y} \le  2k\sqrt{E_{s}}  | +1)
		$}
\end{equation}
To compute this probability, calculate the probability that the real component $y ~(\Re_{y})$ exceeds 0 and does not surpass $2k\sqrt{E_{s}} $, and similarly for the imaginary component $y ~(\Im_{y})$ .
Then integrate the conditional PDF from 0 to $2k\sqrt{E_{s}}$ for both the real and imaginary components.

For instance, to determine the likelihood that the real component $y ~(\Re_{y})$ of the received signal falls within the specified range of 0 to 2, the calculation involves the integration of the PDF across two distinct regions:
1) Probability of exceeding the upper limit: calculate the probability that the real component exceeds the upper limit of 2, extending to infinity $\infty$.
2) Probability of falling below the lower limit: calculate the probability that the real component is less than the lower boundary of 0, extending to negative infinity $-\infty$. 
Since the total probability for any variable within its complete range is 1, the probability that the real component falls within the range from 0 to 2 is obtained by subtracting the sum of the probabilities calculated in steps 1 and 2 from 1. This approach uses the complement rule to derive the probability for the desired range as follows:
\begin{equation}
p(\Re_y \text{ within } 0 \text{ to } 2) = 1 - [p(\Re_y > 2) + p(\Re_y < 0)]
\end{equation}
This method effectively segments the total probability distribution into parts that are outside the interval of interest and subtracts their cumulative impact from unity to find the probability of the variable lying within the specified range.

In general, using the CDF and the error function (erfc), the probability for each component can be derived from:
\begin{equation}
	\resizebox{0.45\textwidth}{!}{$
	\begin{split}
		p(\Re_{y}> 0, \Re_{y} \le  2k\sqrt{E_{s}} | +1) = & 1- \\
		& \big[ \dfrac{1}{\sqrt{\pi N_{0}}} \int_{-\infty}^{0} e^{- \dfrac{(y - k\sqrt{E_{s}})^{2}}{N_{0}}} dy  + \\
		& \dfrac{1}{\sqrt{\pi N_{0}}} \int_{+2k\sqrt{E_{s}}}^{\infty} e^{- \dfrac{(y - k\sqrt{E_{s}})^{2}}{N_{0}}} dy  \big] \\
		= & 1 - erfc\big(k\sqrt{\frac{E_{s}}{N_{0}}}\big)
	\end{split}
	$}
\end{equation}
Similarly,
\begin{equation}
			p(\Im_{y}> 0, \Im_{y} \le  2k\sqrt{E_{s}} | +1) 
			= 1 - erfc\big(k\sqrt{\frac{E_{s}}{N_{0}}}\big)
\end{equation}
The combined probability that the symbol is decoded correctly (both real and imaginary parts fall within the designated range) is:
\begin{equation}
		p(c| inside) =  \big[ 1 - erfc\big(k\sqrt{\frac{E_{s}}{N_{0}}}\big) \big] \big[ 1 - erfc\big(k\sqrt{\frac{E_{s}}{N_{0}}}\big) \big].
\end{equation}
And consequently, the error probability is:
\begin{equation}
	\begin{split}
		p(e| inside) = & 1- \big[ 1 - erfc\big(k\sqrt{\frac{E_{s}}{N_{0}}}\big) \big]^{2} \\
		= &  2erfc\big(k\sqrt{\frac{E_{s}}{N_{0}}}\big)	- erfc^{2}\big(k\sqrt{\frac{E_{s}}{N_{0}}}\big). \\
	\end{split}
\end{equation}

\subsubsection{ For corner symbols in an M-QAM constellation, such as $I = +7, Q= +7$}, the PDF for the received signal $y$, given that this specific symbol ${+7k\sqrt{E_{s}}, +7k\sqrt{E_{s}}}$ was transmitted, is defined as:
\begin{equation}
	p(y | corner) = \dfrac{1}{\sqrt{N_{0}\pi}} e^{- \dfrac{(y - 7k\sqrt{E_{s}})^{2}}{N_{0}}}
\end{equation}
This formula calculates the likelihood of $y$ based on its deviation from the expected position of the corner symbol, adjusted by the noise power $N_0$.
To determine if the corner symbol is correctly decoded, both the real $y ~(\Re_{y})$ and imaginary $y ~(\Im_{y})$ parts of the received symbol must exceed a threshold of 6. This ensures the symbol is decoded as $I = +7, Q= +7$ and not mistaken for a neighboring symbol. The probability of correct decoding is expressed as:
\begin{equation}
	\resizebox{0.45\textwidth}{!}{$
		p(c| corner) =  p(\Re_{y}> 6, \Re_{y} \le  \infty | +7) 
		p(\Im_{y} > 6, \Im_{y} \le  \infty | +7)
		$}
\end{equation}
The thresholds are determined by integrating the tail of the conditional PDF from 6 to $\infty$. This provides the probability that $y$ exceeds the threshold, indicating correct symbol detection:
\begin{equation}
	\resizebox{0.45\textwidth}{!}{$
		\begin{split}
			p(\Re_{y}> 6, \Re_{y} \le  \infty | +7) = & \dfrac{1}{\sqrt{\pi N_{0}}} \int_{6k\sqrt{E_{s}}}^{\infty} e^{- \dfrac{(y - 7k\sqrt{E_{s}})^{2}}{N_{0}}} dy  \\
			= & 1 - \dfrac{1}{2}erfc\big(k\sqrt{\frac{E_{s}}{N_{0}}}\big).
		\end{split}
		$}
\end{equation}
Similarly, for the imaginary part:
\begin{equation}
	p(\Im_{y}> 6, \Im_{y} \le  \infty | +7) 
	= 1 - \dfrac{1}{2}erfc\big(k\sqrt{\frac{E_{s}}{N_{0}}}\big).
\end{equation}
By multiplying the probabilities for the real and imaginary components, the overall probability of correctly decoding the corner symbol is:
\begin{equation}
	p(c| corner) =  \big[ 1 - \dfrac{1}{2}erfc\big(k\sqrt{\frac{E_{s}}{N_{0}}}\big) \big] \big[ 1 - \dfrac{1}{2}erfc\big(k\sqrt{\frac{E_{s}}{N_{0}}}\big) \big].
\end{equation}
The likelihood that the symbol is decoded erroneously (i.e., at least one component is incorrectly decoded) is given by the complement of the correct decoding probability:
\begin{equation}
	\begin{split}
		p(e| corner) = & 1- \big[ 1 - \dfrac{1}{2}erfc\big(k\sqrt{\frac{E_{s}}{N_{0}}}\big) \big]^{2} \\
		= &  erfc\big(k\sqrt{\frac{E_{s}}{N_{0}}}\big)	- \dfrac{1}{4}erfc^{2}\big(k\sqrt{\frac{E_{s}}{N_{0}}}\big). \\
	\end{split}
\end{equation}

\subsubsection{For symbols that are neither at the corners nor strictly inside the grid, such as $I = +7, Q= +1$ }, a detailed error analysis involves assessing the real and imaginary components of the received signal $y$. Correct decoding for these symbols is contingent upon the components falling within specific ranges. The probability of correct demodulation is,
\begin{equation}
	\resizebox{0.45\textwidth}{!}{$
		p(c| side) =  p(\Re_{y}> 6, \Re_{y} \le  \infty | +7) 
		p(\Im_{y} > 0, \Im_{y} \le 2k\sqrt{E_{s}} | +1).
		$}
\end{equation}
Correct decoding of this symbol requires the real part of $y ~(\Re_{y})$ must be greater than 6 and can extend to $\infty$, and the imaginary part of $y ~(\Im_{y})$ must lie between 0 and 2.

The probability that $y ~(\Re_{y})$ falls within the correct range is calculated by integrating the tail of the PDF from 6 to $\infty$:
\begin{equation}
	\resizebox{0.45\textwidth}{!}{$
		\begin{split}
			p(\Re_{y}> 6, \Re_{y} \le  \infty | +7) = & \dfrac{1}{\sqrt{\pi N_{0}}} \int_{6k\sqrt{E_{s}}}^{\infty} e^{- \dfrac{(y - 7k\sqrt{E_{s}})^{2}}{N_{0}}} dy  \\
			= & 1 - \dfrac{1}{2}erfc\big(k\sqrt{\frac{E_{s}}{N_{0}}}\big)
		\end{split}
		$}
\end{equation}
For the imaginary component, the integration considers two segments: 1) From $\infty$ to 0 to find the lower boundary. 2) From $2k\sqrt{E_s}$ to $\infty$ for the upper boundary.
The probability that $y ~(\Im_{y})$ falls within 0 to 2 is:
\begin{equation}
	\resizebox{0.45\textwidth}{!}{$
		\begin{split}
			p(\Im_{y}> 0, \Im_{y} \le  2k\sqrt{E_{s}} | +1) = & 1- \\
			& \big[ \dfrac{1}{\sqrt{\pi N_{0}}} \int_{-\infty}^{0} e^{- \dfrac{(y - k\sqrt{E_{s}})^{2}}{N_{0}}} dy  + \\
			& \dfrac{1}{\sqrt{\pi N_{0}}} \int_{+2k\sqrt{E_{s}}}^{\infty} e^{- \dfrac{(y - k\sqrt{E_{s}})^{2}}{N_{0}}} dy  \big] \\
			= & 1 - erfc\big(k\sqrt{\frac{E_{s}}{N_{0}}}\big).
		\end{split}
		$}
\end{equation}
Combining these probabilities, the overall likelihood that the symbol $I = +7, Q= +1$ is decoded correctly is:
\begin{equation}
	\begin{split}
		p(c| side) =  & \big[ 1 - erfc\big(k\sqrt{\frac{E_{s}}{N_{0}}}\big) \big] \big[ 1 - \dfrac{1}{2}erfc\big(k\sqrt{\frac{E_{s}}{N_{0}}}\big) \big] \\
		= & 1 - \dfrac{3}{2}erfc\big(k\sqrt{\frac{E_{s}}{N_{0}}}\big) + \dfrac{1}{2}erfc^{2}\big(k\sqrt{\frac{E_{s}}{N_{0}}}\big) \\
	\end{split}
\end{equation}

Finally, the probability that the symbol is decoded incorrectly, which occurs if either the real or imaginary part falls outside their respective correct ranges, is:
\begin{equation}
	\begin{split}
		p(e| side) = & 1- p(c| side) \\
		= &  \dfrac{3}{2}erfc\big(k\sqrt{\frac{E_{s}}{N_{0}}}\big) + \dfrac{1}{2}erfc^{2}\big(k\sqrt{\frac{E_{s}}{N_{0}}}\big) \\
	\end{split}
\end{equation} 

\subsection{Symbol Error Probabilities Across Constellation Points}
Given the previously calculated error probabilities for different types of constellation points within an M-QAM system, the total SER is computed by averaging the error probabilities across all constellation point categories. This approach considers the varying likelihood of error occurrence depending on the location of each symbol within the constellation grid.
The joint symbol error rate is calculated using the formula:
\begin{equation}
	\begin{split}
		P(e| MQAM) =  & \dfrac{1}{M}\big[ N_{inside}p(e| inside) + \\
		& N_{corner}p(e| corner) + N_{side}p(e| side) \big].
	\end{split}
\end{equation}
This equation effectively weights the error probabilities by the number of symbols in each category relative to the total number of symbols $M$.
Integrating the individual probabilities:
\begin{equation}
	\begin{split}
		P(e| MQAM)	= &  2(1- \dfrac{1}{\sqrt{M}} )erfc\big(k\sqrt{\frac{E_{s}}{N_{0}}}\big) - \\
		& (1- \dfrac{2}{\sqrt{M}}+\dfrac{1}{M})erfc^{2}\big(k\sqrt{\frac{E_{s}}{N_{0}}}\big). \\
	\end{split}
\end{equation}
Utilizing the $Q-$function to express the error probability provides a more intuitive understanding of the system’s performance under Gaussian noise \cite{10.1002/0471200697.ch8}:
\begin{equation}\label{BER-MQAM-AWGN}
	\begin{split}
		P(e| MQAM)	= &  4(\dfrac{\sqrt{M}-1}{\sqrt{M}} )Q\big(\sqrt{\frac{3E_{s}}{N_{0}(M-1)}}\big) - \\
		& 4(\dfrac{\sqrt{M}-1}{\sqrt{M}})^{2}Q^{2}\big(\sqrt{\frac{3E_{s}}{N_{0}(M-1)}}\big) \\
	\end{split}
\end{equation}
This representation is particularly useful for communications engineers designing systems to meet specific performance criteria under normal operating conditions.
The BER is derived from the SER considering the modulation efficiency.
$E_{s}/N_{0} = qE_{b}/N_{0}$, $q = log_{2}(M)$, the BER = $\frac{1}{log_2(M)}$SER.
For a 16-QAM system example, SER is calculated as \cite{Oyetola2020SYMBOLEP}:
\begin{equation}
	\begin{split}
		P_{s}(e| 16QAM)	= &  \dfrac{3}{2}erfc\big(\sqrt{\frac{E_{s}}{10N_{0}}}\big) - 
		\dfrac{9}{16}erfc^{2}\big(\sqrt{\frac{E_{s}}{10N_{0}}}\big) \\
		\approx & \dfrac{3}{2}erfc\big(\sqrt{\frac{E_{s}}{10N_{0}}}\big)
	\end{split}
\end{equation}
The corresponding BER for Gray coded 16-QAM in an AWGN environment where $q = 4$ is:
 \begin{equation}
	P_{b}(e| 16QAM)	= \dfrac{3}{8}erfc\big(\sqrt{\frac{2E_{b}}{5N_{0}}}\big)
\end{equation}

\section{The Calculation of BER for M-QAM Modulation Scheme in Rayleigh Fading Channel}

To get the probability of error over Rayleigh fading, we should average the BER in AWGN over the fading channel as follow \cite{soton258001}:
\begin{equation}\label{general_BER_fading}
	P_{b} =  \int_{0}^{\infty}  P_{e}(\gamma)p_{\gamma}(\gamma) d\gamma,
\end{equation}
where the probability density function of $\gamma$ is,
\begin{equation}
	\begin{split}
		p_{\gamma}(\gamma) = & \frac{1}{E_{b}/N_{0}}e^{\frac{-\gamma}{E_{b}/N_{0}}}, ~ \gamma \ge 0 \\
		= & \frac{1}{\bar{\gamma}}e^{\frac{-\gamma}{\bar{\gamma}}}, \\
	\end{split}
\end{equation}
and $\bar{\gamma} = E_{b}/N_{0}$.
WE will use the following Craig forms for the Q-function.
\begin{equation}\label{Q_creig}
	Q(x) =  \dfrac{1}{\pi}\int_{0}^{\frac{\pi}{2}} e^{-\dfrac{x^{2}}{2 \sin^{2} (\theta)}}  d \theta,
\end{equation}
and
\begin{equation}\label{Q_square}
	Q^{2}(x) =  \dfrac{1}{\pi}\int_{0}^{\frac{\pi}{4}} e^{-\dfrac{x^{2}}{2 \sin^{2} (\theta)}}  d \theta.
\end{equation}
By replacing $x^{2} = \dfrac{3\gamma}{M-1}$ in (\ref{Q_creig}) and (\ref{Q_square}). Then by substituting them in (\ref{BER-MQAM-AWGN}) and (\ref{general_BER_fading}), the probability of error in Rayleigh fading will be
\begin{equation}
		\begin{split}
		P_{e-MQAM} 	= &  \dfrac{4}{\pi}(\dfrac{\sqrt{M}-1}{\sqrt{M}}) \int_{0}^{\frac{\pi}{2}} \int_{0}^{\infty} e^{- \frac{3}{M-1} \frac{\gamma}{2 \sin^{2} (\theta)}}  d\theta p_{\gamma}(\gamma) d\gamma - \\
		& \dfrac{4}{\pi}(\dfrac{\sqrt{M}-1}{\sqrt{M}})^{2}  \int_{0}^{\frac{\pi}{4}}  \int_{0}^{\infty} e^{-\frac{3}{M-1} \frac{\gamma}{2 \sin^{2} (\theta)}}  d \theta p_{\gamma}(\gamma) d\gamma.\\
	\end{split}
\end{equation}
To solve this we will use the Laplace transformation which implies:
\begin{equation}\label{Laplace}
	\begin{split}
	F_{\gamma}(S) = & \int_{0}^{\infty} e^{-S\gamma} p_{\gamma}(\gamma) d \gamma \\
	= & \int_{0}^{\infty} e^{-S\gamma}  \frac{1}{\bar{\gamma}}e^{\frac{-\gamma}{\bar{\gamma}}} d \gamma \\
	= & \dfrac{1}{1 + S\bar{\gamma}}.
	\end{split}
\end{equation}
Let $S = (\frac{3}{M-1}) \frac{1}{2 \sin^{2} (\theta)}$.
\begin{equation}\label{Pmqam_two_parts}
	\begin{split}
		P_{e-MQAM} 	= &  \dfrac{4}{\pi}(\dfrac{\sqrt{M}-1}{\sqrt{M}}) \int_{0}^{\frac{\pi}{2}}  F_{\gamma}\big( (\frac{3}{M-1}) \frac{1}{2 \sin^{2} (\theta)} \big) d\theta - \\
		& \dfrac{4}{\pi}(\dfrac{\sqrt{M}-1}{\sqrt{M}})^{2}  \int_{0}^{\frac{\pi}{4}} F_{\gamma}\big( (\frac{3}{M-1}) \frac{1}{2 \sin^{2} (\theta)} \big)   d\theta \\
		= &  \dfrac{4}{\pi}(\dfrac{\sqrt{M}-1}{\sqrt{M}}) \int_{0}^{\frac{\pi}{2}}  \big( \dfrac{1}{1 + (\frac{3}{M-1}) \frac{\bar{\gamma}}{2 \sin^{2} (\theta)}} \big) d\theta - \\
		& \dfrac{4}{\pi}(\dfrac{\sqrt{M}-1}{\sqrt{M}})^{2}  \int_{0}^{\frac{\pi}{4}} \big( \dfrac{1}{1 + (\frac{3}{M-1}) \frac{\bar{\gamma}}{2 \sin^{2} (\theta)}} \big)   d\theta \\
		= & P_{e-MQAM} (1) - P_{e-MQAM} (2).
	\end{split}
\end{equation}
Let's solve the first part of the previous equation:
\begin{equation}
	\begin{split}
		P_{e-MQAM}(1) 	= &  \dfrac{4}{\pi}(\dfrac{\sqrt{M}-1}{\sqrt{M}}) \int_{0}^{\frac{\pi}{2}}  \big(1 + (\frac{3}{M-1}) \frac{\bar{\gamma}}{2 \sin^{2} (\theta)}\big)^{-1} d\theta \\
		= & \dfrac{4}{\pi}(\dfrac{\sqrt{M}-1}{\sqrt{M}}) \int_{0}^{\frac{\pi}{2}}  \big(1-1 + \dfrac{2 \sin^{2} (\theta)}{2 \sin^{2} (\theta) + (\frac{3}{M-1})\bar{\gamma}} \big) d\theta \\
		= & \dfrac{4}{\pi}(\dfrac{\sqrt{M}-1}{\sqrt{M}}) \int_{0}^{\frac{\pi}{2}}  \big(1- \dfrac{2 \sin^{2} (\theta) + (\frac{3}{M-1})\bar{\gamma}}{2 \sin^{2} (\theta) + (\frac{3}{M-1})\bar{\gamma}} +\\ 
		 & \dfrac{2 \sin^{2} (\theta)}{2 \sin^{2} (\theta) + (\frac{3}{M-1})\bar{\gamma}} \big) d\theta \\
		 = & 2(\dfrac{\sqrt{M}-1}{\sqrt{M}}) - \\ 
		 & \dfrac{4}{\pi}(\dfrac{\sqrt{M}-1}{\sqrt{M}}) \int_{0}^{\frac{\pi}{2}}  \big( 
		  \dfrac{(\frac{3}{M-1})\bar{\gamma}}{2 \sin^{2} (\theta) + (\frac{3}{M-1})\bar{\gamma}} \big) d\theta. \\
	\end{split}
\end{equation}
Now, let $\tan (\theta) = t$, $\dfrac{dt}{d\theta} = \sec^{2}(\theta) = 1 + \tan^{2}(\theta) = 1 + t^{2}$, $d\theta = \dfrac{1}{1 + t^{2}}dt$, and $\sin(\theta) = \dfrac{t}{\sqrt{1 + t^{2}}}$.
\begin{equation}
	\begin{split}
		P_{e-MQAM}(1) = & 2(\dfrac{\sqrt{M}-1}{\sqrt{M}}) - \\ 
		& \dfrac{4}{\pi}(\dfrac{\sqrt{M}-1}{\sqrt{M}}) \int_{\tan(0)}^{\tan(\frac{\pi}{2})}  \big( \dfrac{(\frac{3}{M-1})\bar{\gamma}}{2 \dfrac{t^{2}}{1 + t^{2}} + (\frac{3}{M-1})\bar{\gamma}} \big) \\
		& * \dfrac{1}{1 + t^{2}}dt \\
		= & 2(\dfrac{\sqrt{M}-1}{\sqrt{M}}) - \\ 
		& \dfrac{4}{\pi}(\dfrac{\sqrt{M}-1}{\sqrt{M}}) (\frac{3}{M-1})\bar{\gamma} \\
		&* \int_{\tan(0)}^{\tan(\frac{\pi}{2})}  \big( \dfrac{1 + t^{2}}{2 t^{2} + (\frac{3}{M-1})\bar{\gamma}(1 + t^{2})} \big) \\
		& * \dfrac{1}{1 + t^{2}}dt \\
		= & 2(\dfrac{\sqrt{M}-1}{\sqrt{M}}) - \\ 
		& \dfrac{4}{\pi}(\dfrac{\sqrt{M}-1}{\sqrt{M}}) (\frac{3}{M-1})\bar{\gamma} \\
		&* \int_{\tan(0)}^{\tan(\frac{\pi}{2})}  \big( \dfrac{1}{(\frac{3}{M-1})\bar{\gamma} + (2+(\frac{3}{M-1})\bar{\gamma}) t^{2}} \big) dt. \\
	\end{split}
\end{equation}
To solve the integral, let $a =( \frac{3}{M-1})\bar{\gamma}$ and $b = 2 + (\frac{3}{M-1})\bar{\gamma}$ \cite{Oyetola2020SYMBOLEP}. Then the integral will be:
\begin{equation}\label{MQAM_BER_1st_integral}
	\begin{split}
 	\int_{\tan(0)}^{\tan(\frac{\pi}{2})}  \big( \dfrac{1}{(\frac{3}{M-1})\bar{\gamma} + (2+(\frac{3}{M-1})\bar{\gamma}) t^{2}} \big) dt = & \\
 	& \int_{\tan(0)}^{\tan(\frac{\pi}{2})} \dfrac{1}{a+ b t^{2}}dt\\
	\end{split}
\end{equation}
To solve the indefinite integral $\int \dfrac{1}{a+ b t^{2}}dt$, let $\tan(\theta)  = \frac{\sqrt{b}}{\sqrt{a}}x$, $x = \sqrt{\frac{a}{b}} \tan (\theta)$, $dx = \sqrt{\frac{a}{b}}\frac{1}{\cos^{2}(\theta)}d\theta$, $\theta = \tan^{-1}(\sqrt{\frac{b}{a}}x)$, $\cos (\theta) = \frac{\sqrt{a}}{\sqrt{a + b x^{2}}}$, and $a + b x^{2} = \frac{a}{\cos (\theta)}$.
\begin{equation}\label{indefinite_integral}
	\begin{split}
		\int \dfrac{1}{a+ b t^{2}}dt =	& \sqrt{\frac{a}{b}}\int \dfrac{\dfrac{1}{\cos^{2}(\theta)}}{\dfrac{a}{\cos^{2}(\theta)}}d\theta \\
		=	& \dfrac{1}{a}\sqrt{\frac{a}{b}}\int d\theta \\
		=	& \dfrac{1}{a}\sqrt{\frac{a}{b}} \tan^{-1}(\sqrt{\frac{b}{a}}x). \\
	\end{split}
\end{equation}
An important assumption has been made as $\sqrt{\frac{b}{a}} \approx 1$, then $\tan (\tan^{-1}(\theta)) = \theta$. By applying the last result to (\ref{MQAM_BER_1st_integral}), we get
\begin{equation}
	\begin{split}
		\int_{\tan(0)}^{\tan(\frac{\pi}{2})} \dfrac{1}{a+ b t^{2}}dt = & \dfrac{1}{a}\sqrt{\frac{a}{b}} \dfrac{\pi}{2} \\
		= & \dfrac{1}{(\frac{3}{M-1})\bar{\gamma}}\sqrt{\frac{(\frac{3}{M-1})\bar{\gamma}}{2 + (\frac{3}{M-1})\bar{\gamma}}} \dfrac{\pi}{2}
	\end{split}
\end{equation}
\begin{equation}
	\begin{split}
		P_{e-MQAM}(1) = & 2(\dfrac{\sqrt{M}-1}{\sqrt{M}}) \big[ 1  - \\ 
		& \sqrt{\frac{(\frac{3}{M-1})\bar{\gamma}}{2 + (\frac{3}{M-1})\bar{\gamma}}} \big]
	\end{split}
\end{equation}
Let's solve the second part of (\ref{Pmqam_two_parts}) which can reduced in similar to the first part as follow:
\begin{equation}
	\begin{split}
		P_{e-MQAM} (2)	=
		& \dfrac{4}{\pi}(\dfrac{\sqrt{M}-1}{\sqrt{M}})^{2} \int_{0}^{\frac{\pi}{4}} \big( \dfrac{1}{1 + (\frac{3}{M-1}) \frac{\bar{\gamma}}{2 \sin^{2} (\theta)}} \big)   d\theta \\
		= & (\dfrac{\sqrt{M}-1}{\sqrt{M}})^{2}  - \\ 
		& \dfrac{4}{\pi}(\dfrac{\sqrt{M}-1}{\sqrt{M}})^{2} \int_{0}^{\frac{\pi}{4}}  \big( 
		\dfrac{(\frac{3}{M-1})\bar{\gamma}}{2 \sin^{2} (\theta) + (\frac{3}{M-1})\bar{\gamma}} \big) d\theta. \\
	\end{split}
\end{equation}
let $\tan (\theta) = t$, $\dfrac{dt}{d\theta} = \sec^{2}(\theta) = 1 + \tan^{2}(\theta) = 1 + t^{2}$, $d\theta = \dfrac{1}{1 + t^{2}}dt$, and $\sin(\theta) = \dfrac{t}{\sqrt{1 + t^{2}}}$.
\begin{equation}
	\begin{split}
		P_{e-MQAM}(2) = & (\dfrac{\sqrt{M}-1}{\sqrt{M}})^{2} - \\ 
		& \dfrac{4}{\pi}(\dfrac{\sqrt{M}-1}{\sqrt{M}})^{2} (\frac{3}{M-1})\bar{\gamma} \\
		&* \int_{\tan(0)}^{\tan(\frac{\pi}{4})}  \big( \dfrac{1}{(\frac{3}{M-1})\bar{\gamma} + (2+(\frac{3}{M-1})\bar{\gamma}) t^{2}} \big) dt. \\
	\end{split}
\end{equation}
Using the solution of the indefinite integral $\int \dfrac{1}{a+ b t^{2}}dt$ in (\ref{indefinite_integral}) and replace $t = \tan(\theta)$, we get:
\begin{equation}
	\begin{split}
		\int_{\tan(0)}^{\tan(\frac{\pi}{4})} \dfrac{1}{a+ b t^{2}}dt = & \dfrac{1}{a}\sqrt{\frac{a}{b}} \tan^{-1}(\sqrt{\frac{b}{a}}\tan(\theta))\big|_{0}^{\frac{\pi}{4}} \\
		= & \dfrac{1}{(\frac{3}{M-1})\bar{\gamma}}\sqrt{\frac{(\frac{3}{M-1})\bar{\gamma}}{2 + (\frac{3}{M-1})\bar{\gamma}}} \\
		& * \tan^{-1} \big(  \sqrt{\frac{2 + (\frac{3}{M-1})\bar{\gamma}}{(\frac{3}{M-1})\bar{\gamma}}}\big)
	\end{split}
\end{equation}
Therefore,
\begin{equation}
	\begin{split}
		P_{e-MQAM}(2) = & (\dfrac{\sqrt{M}-1}{\sqrt{M}})^{2} \big[ 1 - \sqrt{\frac{(\frac{3}{M-1})\bar{\gamma}}{2 + (\frac{3}{M-1})\bar{\gamma}}} \\
		& * \dfrac{4}{\pi}\tan^{-1} \big(  \sqrt{\frac{2 + (\frac{3}{M-1})\bar{\gamma}}{(\frac{3}{M-1})\bar{\gamma}}}\big) \big] \\
		= & (\dfrac{\sqrt{M}-1}{\sqrt{M}})^{2} \big[ 1 - \sqrt{\frac{1.5\bar{\gamma}}{M-1 + 1.5\bar{\gamma}}} \\
		& * \dfrac{4}{\pi}\tan^{-1} \big(  \sqrt{\frac{M-1 + 1.5\bar{\gamma}}{1.5\bar{\gamma}}}\big) \big] \\
	\end{split}
\end{equation}
The total symbol error probability for M-QAM is:
\begin{equation}
	\begin{split}
		P_{e-MQAM}	= & 2(\dfrac{\sqrt{M}-1}{\sqrt{M}}) \big[ 1  -
		 \sqrt{\frac{1.5\bar{\gamma}}{M-1 + 1.5\bar{\gamma}}} \big]- \\
		& (\dfrac{\sqrt{M}-1}{\sqrt{M}})^{2} \big[ 1 - \sqrt{\frac{1.5\bar{\gamma}}{M-1 + 1.5\bar{\gamma}}} \\
		& * \dfrac{4}{\pi}\tan^{-1} \big(  \sqrt{\frac{M-1 + 1.5\bar{\gamma}}{1.5\bar{\gamma}}}\big) \big] \\
	\end{split}
\end{equation}

\section{Conclusion}
In this paper, we have presented a unified approach for deriving the probability of error formulations applicable to BPSK, 16-QAM, and 64-QAM in Rayleigh fading channels. By considering the statistical properties of Rayleigh fading and the modulation characteristics of each scheme, our approach provides a comprehensive framework for analyzing error performance across different modulation schemes.
Importantly, our approach applies to solving all types of integration processes, producing accurate solutions that account for the complexities of Rayleigh fading and modulation schemes. By establishing a unified framework, we have simplified the analysis process and facilitated a deeper understanding of error behavior, enabling researchers, engineers, and practitioners to design and optimize communication systems with confidence.
Looking ahead, further research can explore additional modulation schemes, channel models, and advanced signal processing techniques to extend the applicability and effectiveness of our approach. The findings presented herein serve as a foundation for future research and development efforts aimed at achieving reliable communication in challenging fading environ

\appendices
\section{Common Used Functions}\label{Common_used_Probabilities}
The complimentary distribution function (CDF) is calculated by:
\begin{equation}\label{CDF}
	F_{x}(x) = \dfrac{1}{\sqrt{2 \pi \sigma^{2}}} \int_{-\infty}^{x} e^{\dfrac{-(\alpha - \mu)^{2}}{2\sigma^{2}}} d\alpha
\end{equation}
The complimentary error function is:
\begin{equation}\label{erfc-function}
	\begin{split}
		erfc\big(x\big) = & \dfrac{2}{\sqrt{\pi }} \int_{x}^{\infty} e^{-t^{2}} dt \\
		= & 1 - erf(x),\\
	\end{split}
\end{equation}
and error $erf(x)$ function is expressed by:
\begin{equation}\label{erf-function}
		erf\big(x\big) = \dfrac{2}{\sqrt{\pi }} \int_{0}^{x} e^{-t^{2}} dt,
\end{equation}
the derivative of $erf(x)$ function is:
\begin{equation}\label{derivative_erf-function}
		\dfrac{d}{dx} erf\big(x\big) = -\dfrac{2}{\sqrt{\pi }} e^{-x^{2}}
\end{equation}
And also the definite integral of $erfc(x)$, simply calculating it by parts ($\int u dv = uv - \int v du$). By taking $u = erfc(x)$, $du = \dfrac{2}{\sqrt{\pi}} e^{-t^{2}} $, $dv = dt$, and $v = x$. 
\begin{equation}
	\int erfc\big(x\big)dx = xerfc(x) - \dfrac{2}{\sqrt{\pi}} \int xe^{-x^{2}} dx,
\end{equation}
By taking $u =- x^{2}$ and $du = -2xdx$.
\begin{equation}\label{integral_erfc-function}
	\begin{split}
			-\dfrac{2}{\sqrt{\pi}} \int xe^{-x^{2}} dx = & \dfrac{1}{\sqrt{\pi}} \int e^{u} du, \\
			= &  \dfrac{1}{\sqrt{\pi}} e^{u} + C\\
			= &  \dfrac{1}{\sqrt{\pi}} e^{-x^{2}} + C\\
	\end{split}
\end{equation}
\begin{equation}
	\int erfc\big(x\big)dx = xerfc(x) + \dfrac{1}{\sqrt{\pi}} e^{-x^{2}} + C
\end{equation}

\section{The Relation Between The $Q(x)$ and $erfc(x)$ Functions}
The Q-function CDF is:
\begin{equation}\label{Q-function}
	\begin{split}
		Q(x) = & Pr [X > x] = 1 - F(x)\\
		= & \dfrac{1}{\sqrt{2 \pi}} \int_{x}^{\infty} e^{\dfrac{-u^{2}}{2}} du \\
		= & \dfrac{1}{2} erfc\big(\sqrt{\frac{x}{2}}\big)\\
		= & \dfrac{1}{2} [1 - erf\big(\sqrt{\frac{x}{2}}\big)], \\
	\end{split}
\end{equation}
By taking $z^{2} =\dfrac{u^{2}}{2}$, $z = \dfrac{u}{\sqrt{2}}$, $dz = \dfrac{1}{\sqrt{2}}du$, and $du = \sqrt{2}dz$. The integration limits will be as follows: $u = x \Rightarrow z = \dfrac{ x}{\sqrt{2}}$ and $u = \infty \Rightarrow z = \infty$.
\begin{equation}\label{Q_to_erfc}
	\begin{split}
		Q(x)
		= & \dfrac{\sqrt{\pi}}{\sqrt{2 \pi}} \int_{\frac{ x}{\sqrt{2}}}^{\infty} e^{-z^{2}} dz \\
		= & \dfrac{1}{2} erfc\big(\frac{x}{\sqrt{2}}\big)
	\end{split}
\end{equation}
To get the Q-function from erfc function, we can follow the simple below calculations:
\begin{equation}\label{}
	\begin{split}
		erfc\big(x\big) = & \dfrac{2}{\sqrt{\pi }} \int_{x}^{\infty} e^{-t^{2}} dt \\
		= & 1 - erf(x),\\
	\end{split}
\end{equation}
By taking $\dfrac{z^{2}}{2} = u^{2}$, $z = u\sqrt{2}$, $dz = \sqrt{2}du$, and $du = \dfrac{1}{\sqrt{2}}dz$. The integration limits will be as follows: $u = x \Rightarrow z =  x\sqrt{2}$ and $u = \infty \Rightarrow z = \infty$.
\begin{equation}\label{}
	\begin{split}
		erfc\big(x\big) = & \dfrac{2}{\sqrt{\pi }} \int_{x}^{\infty} e^{-u^{2}} du \\
		= & \dfrac{2}{\sqrt{2\pi }} \int_{\sqrt{2}x}^{\infty} e^{-\frac{z^{2}}{2}} dz,\\
		= & 2 Q(x\sqrt{2}),\\
	\end{split}
\end{equation}

\section*{Acknowledgment}
The authors wish to thank Krishna Sankar who is Principal Systems Design Engineer of the HMicro, Bangalore area, India for the informative webpage on wireless communications and simplified explanation of BER of BPSK, PAM, and QAM in AWGN and Rayleigh fading channels \cite{dsplog}.
\bibliographystyle{IEEEtran}
\bibliography{IEEEabrv,main}

\end{document}